\documentstyle[aps,prd,preprint,amstex,amssymb,tighten,eqsecnum,draft]{revtex}
\makeatletter
\input epsf
\input psfrag.sty
\makeatother
\newcommand{\be}{\begin{equation}}
\newcommand{\ee}{\end{equation}}
\def\cf{cf.\hbox{}}
\def\eg{e.g.\hbox{}}

\def\etal{{\it et~al.\/}}
\def\ie{i.e.\hbox{}}

\def\citeprefix{Ref.~}
\def\citesprefix{Refs.~}
\def\BH{{\scriptscriptstyle\text{BH}}}
\def\constant{{\text{constant}}}

\def\fit{{\text{fit}}}

\def\init{{\text{init}}}

\def\max{{\text{max}}}          
\def\MS{{\scriptscriptstyle\text{MS}}}
\def\outer{{\text{outer}}}
\def\pred{{\text{pred}}}
\def\total{{\text{total}}}
\def\ccbeta{\eta}
\def\Vr{V/r}                    
\def\parVr{(V/r)}               
\def\parVrp{{(V/r)'}}           

\def\f{{\sf f}}
\def\y{{\sf y}}
\def\csmash#1{{\hbox to 0em{\hss{#1}\hss}}}     

\def\tfrac#1#2{{\textstyle \frac{#1}{#2}}}
\def\half{\frac{1}{2}}
\def\thalf{\tfrac{1}{2}}

\def\gtsim{\gtrsim}
\def\ltsim{\lesssim}
\def\Bizon{{\text{Bizo\'{n}}}}

\def\Gomez{{\text{G\'{o}mez}}}

\def\Purrer{{\text{P\"urrer}}}

\begin{document}
\preprint{UWThPh-2000-12}
\title{Type~II Critical Collapse of a Self-Gravitating
       Nonlinear $\sigma$-Model}
\author{Sascha Husa$^{(2,3)}$,
        Christiane Lechner$^{(1)}$, Michael \Purrer$^{(1)}$,            \\
        Jonathan Thornburg$^{(1)}$, and Peter C.~Aichelburg$^{(1)}$}
\address{$^{(1)}$
         Institut f\"{u}r Theoretische Physik,
         Universit\"{a}t Wien,                                          \\
         Boltzmanngasse~5, A-1090 Wien, Austria                         \\
         $^{(2)}$
         Department of Physics and Astronomy,
         University of Pittsburgh,                                      \\
         3941 O'Hara Street,
         Pittsburgh PA 15260, USA                                       \\
         $^{(3)}$
         Max-Planck-Institut f\"ur Gravitationsphysik,
         Albert-Einstein-Institut\\
         Am M\"uhlenberg 1, D-14476 Golm, Germany}
\date{\today}
\maketitle


\begin{abstract}
We report on the existence and phenomenology of type~II critical
collapse within the one-parameter family of SU(2) $\sigma$-models
coupled to gravity.  Numerical investigations in spherical
symmetry show discretely self-similar (DSS) behavior
at the threshold of black hole formation for values of the dimensionless
coupling constant $\ccbeta$ ranging from 0.2 to 100; at 0.18 we see small
deviations from DSS.
While the echoing period
$\Delta$ of the critical solution rises sharply towards the lower
limit of this range, the characteristic mass scaling has a critical
exponent $\gamma$ which is almost independent of $\ccbeta$,
asymptoting to $0.1185 \pm 0.0005$ at large $\ccbeta$.
We also find critical scaling of the scalar
curvature for near-critical initial data.
Our numerical results are based on an outgoing--null-cone formulation
of the Einstein-matter equations, specialized to spherical symmetry.
Our numerically computed initial-data critical parameters $p^*$ show
2nd~order convergence with the grid resolution, and after compensating
for this variation in $p^*$, our individual evolutions are uniformly
2nd~order convergent even very close to criticality.
\end{abstract}


\draft  

\section{Introduction}

Since the numerical investigation of dynamical behavior of a massless
scalar field under the influence of its gravitational forces by
Choptuik (\citeprefix\cite{Choptuik-1993-self-similarity}),
critical behavior has been observed in a number of different
matter models coupled to gravity.  In the context of type II critical
collapse, these models have in common that at
the threshold of black hole formation their dynamics show a universal
characteristic approach to either a discretely (DSS) or continuously (CSS)
self-similar solution.

Nonlinear $\sigma$-fields provide particularly interesting models to study
the dynamics of gravitating self-interacting matter in general relativity.
Besides their applications in physics 
(see \eg{}~\citeprefix\cite{Misner-1978-harmonic-maps}), 
they have a simple
geometrical interpretation as harmonic maps, which have been extensively
studied in the mathematical literature
(see \eg{}~\citesprefix\cite{Eells-Lemaire-1978,Eells-Lemaire-1988}).

Recently \Bizon{}~\etal{}
(\citesprefix\cite{Bizon-1999-existence-of-self-similar-sigma-CSS-solutions,
Bizon-Chmaj-Tabor-1999-sigma-3+1-evolution})
and also independently Liebling~\etal{}
(\citeprefix\cite{Liebling-Hirschmann-Isenberg-1999-sigma-critical})
have observed
critical (threshold) behavior for non-gravitating systems:
The transition between
globally regular time evolution and singularity formation for the
SU(2) $\sigma$--model on Minkowski background. It was shown by \Bizon{}
(\citeprefix\cite{Bizon-1999-existence-of-self-similar-sigma-CSS-solutions})
that this system admits a countably infinite family of
CSS solutions. 
The stable ground state is the endpoint of
singular evolution for supercritical initial data, while the first
excitation, which has one unstable mode, plays the role of the critical
(CSS) solution.

The interesting question arises of what happens if gravity is added
to this system.  The gravitating SU(2) $\sigma$-model is a family of
theories with a {\em dimensionless parameter\/} $\ccbeta$, which acts
as a coupling constant
(for $\ccbeta = 0$ gravity decouples from the
field). 
It was argued in
\citeprefix\cite{Bizon-Chmaj-Tabor-1999-sigma-3+1-evolution}
that the singularity formation in flat space might not be relevant
for black hole formation when gravity is active, since the CSS blowup
excludes the concentration of energy at the singularity.  Since no
asymptotically flat solitonic configurations exist
(\citeprefix\cite{Heusler-1996-No-hair-Theorems}), this suggests
that the only alternative to dispersion or collapse to a black hole
is the formation of a naked singularity.  
Here we
focus on critical phenomena at the threshold of
black hole formation.
As \Bizon{}~\etal{} have pointed out
(\citeprefix\cite{Bizon-Chmaj-Tabor-1999-skyrme-3+1-crit-collapse}),
criticality is expected to depend on the coupling constant $\ccbeta$.
If so, does the system show discrete or continuous self-similarity?
And in which way do critical phenomena depend on the coupling?

In this paper we present
results from a numerical study of the dynamical evolution for the
SU(2) nonlinear $\sigma$-model coupled to gravity in spherical symmetry.
Our code uses a characteristic formulation, specialized to
the spherical symmetry. Initial data are
specified on an outgoing null cone with vertex at the center of
symmetry. The discretized field equations are used to evolve the
matter field and the geometry to future outgoing null cones,
using a nonuniformly spaced set of grid points which follow ingoing
null geodesics.

We find critical behavior at the boundary between black hole formation
and dispersion for values of the coupling constant $\ccbeta$ in the
range of \hbox{0.18--100}.  The critical solution is DSS,
with the echoing period $\Delta$ strongly depending on
$\ccbeta$: As $\ccbeta$ tends to 0.18 from above, $\Delta$ rises sharply.
Moreover we observe small deviations from exact DSS at this smallest
$\ccbeta$ value.  This leads us to conjecture that DSS ceases
to be a critical solution for still smaller values of the coupling
constant.

The organization of this paper is as follows: In
section~\ref{sect-SU(2)-sigma-model} we review the basic properties
of the SU(2) $\sigma$-model in spherical symmetry and discuss the
system of field equations.  We present our main physical results
in section~\ref{sect-results}, and end the main body of the paper
with some conclusions in section~\ref{sect-conclusions}.
In appendix~\ref{app-numerical} we discuss our numerical methods,
which are based on previous work of Goldwirth and Piran
(\citesprefix\cite{Goldwirth-Piran-1987-sssf-2+2,
Goldwirth-Ori-Piran-1989-in-Frontiers}),
Garfinkle (\citeprefix\cite{Garfinkle-1995-sssf-2+2-self-similarity}),
and \Gomez{} and Winicour
(\citesprefix\cite{GLPW-1996-sssf-3+1-and-2+2,
Gomez-Winicour-1992-in-dInverno,
Gomez-Winicour-1992-sssf-2+2-asymptotics,
Gomez-Winicour-1992-sssf-2+2-numerical-methods}).  Finally, in
appendix~\ref{app-convergence} we discuss the convergence of our
numerical evolutions to the continuum limit as the grid resolution
is increased, including both uniform convergence of accuracy
diagnostics within a single evolution, and also convergence of
the numerically computed critical parameter $p^*$ itself.

Conventions are chosen as follows:
spacetime indices are Greek letters, SU(2) indices are uppercase Latin
letters,
the spacetime signature is $(-,+,+,+)$, the Ricci tensor is defined
as $R_{\mu\nu} = {R_{\mu\lambda\nu}}^{\lambda}$
with the sign convention of \citeprefix\cite{Wald},
and the speed of light is set to unity, $c=1$.


\section{The SU(2) $\sigma$-Model in Spherical Symmetry}
\label{sect-SU(2)-sigma-model}

Nonlinear $ \sigma $-models are special cases of harmonic maps from a spacetime
$ ({\mathbf{M}},g_{\mu \nu }) $ into some target manifold $ ({\mathbf{N}},G_{AB}) $
(see, \eg{}, \citeprefix\cite{Misner-1978-harmonic-maps}).
Harmonic maps $ X^{A}(x^{\mu }) $ are defined as the extrema of the simple
geometric action
\begin{equation}
\label{SM}
S = -\frac{\f_{\pi}^2}{2} \int _{{\mathbf{M}}}d^4x \sqrt{|g|}
\, g^{\mu \nu } \partial _{\mu }X^{A}\partial _{\nu }X^{B}
\, G_{AB}(X)
                                                                \, \text{.}
\end{equation}

If the spacetime metric is dynamically coupled to the matter fields $X^A$,
then~\eqref{SM} must be supplemented by the Einstein-Hilbert action.

Variation of the total action with respect to the $\sigma$ field $X^A$ and the
metric $g_{\mu\nu}$ yields the coupled Einstein-$\sigma$ field equations.
The stress-energy tensor resulting from~\eqref{SM} obeys the weak, strong
and dominant energy conditions (\citeprefix\cite{Hawking-Ellis}).
The coupling constant $f_\pi^2$ and the gravitational constant~$G$
enter the equations only in the dimensionless product
$\ccbeta \equiv 4\pi G f_{\pi}^2$, thereby defining a one-parameter family of
distinct gravitating matter models.  The field equations are scale invariant.

For the SU(2) $ \sigma  $-model, the target manifold is taken as $ S^{3} $
with $ G_{AB} $ the ``round'' metric of constant curvature. 
Note that the coupling $\eta$ may be interpreted as
the inverse of the scalar curvature of the target manifold.
In the limit $\eta \to \infty$ our model thus corresponds
to the $\sigma$-model with 3-dimensional flat target manifold.
(This is also easily checked by rescaling the field 
$\phi \to \phi/\sqrt{\eta}$ and performing the limit $\eta \to \infty$ in
Eqs \ref{eqn-phi} and \ref{eqn-hypersurface} -- \ref{eqn-Ethth-redundant}.) 
We restrict
ourselves
to spherically symmetric harmonic maps coupled to gravity, which implies that
the base space (spacetime) must share this symmetry.

We introduce a Bondi coordinate system $ \{u,r,\theta ,\varphi \} $
on spacetime
based upon outgoing null hypersurfaces $ u=\constant $, with the line element
\begin{equation}
\label{metrBondi}
ds^{2}=-e^{2\beta (u,r)}du\left( \frac{V(u,r)}{r}du+2dr\right) +
r^{2}(d\theta ^{2}+\sin ^{2}\theta d\varphi ^{2}),
\end{equation}
and assume that spacetime admits a regular center $r=0$ of spherical
symmetry.  This requires the metric functions near the origin to behave
at fixed retarded time $ u_{0} $ like
\begin{eqnarray}
\beta (u_{0},r) & = & O(r^{2}) \, ,\\
V(u_{0},r) & = & r+O(r^{3}) \, .
\end{eqnarray}
where the guage has been fixed such that
the family of outgoing null cones emanating from the center
is parametrized by the proper time $u$ at the center.
Radial ingoing null geodesics are obtained by integrating the equation
\begin{equation}\label{eqn-geodesic}
\frac{d}{du}r(u) = - \frac{V \big( u, r(u) \big)}{2 r(u)}
                                                                \, \text{.}
\end{equation}

In spherical symmetry the null expansions $ \Theta _{\pm } $ of
inward and outward directed null rays
emanating from $ r=\constant$ surfaces, can be defined as
$ \Theta _{\pm }=2({\mathcal{L}}_{\pm }r)/r $,
where $ {\mathcal{L}}_{\pm } $ is the Lie-derivative along the null directions
$ l_{+}=e^{-2\beta }\partial _{r} $ and
$ l_{-}=2\partial _{u}-(V/r)\partial _{r} $.
Thus we have
\begin{equation}\label{Theta-pm}
\Theta _{+}=\frac{2}{r}e^{-2\beta }
                                                                \, \text{,}
        \qquad
\Theta _{-}=-\frac{2}{r}(\frac{V}{r})
                                                                \, \text{.}
\end{equation}
 Whenever $ \Theta _{+} $ vanishes on some 2-sphere $ r=\constant $, this sphere
is marginally outer trapped. Since this means diverging $ \beta $, the
Bondi-like coordinate system (\ref{metrBondi}) cannot penetrate
a marginally outer trapped surface -- in particular an apparent
horizon.

We introduce polar coordinates $ (\phi ,\Theta ,\Phi ) $ on the target
manifold $ (S^{3},G) $ , and write the SU(2) line element as
\begin{equation}
\label{targetMf-metric}
ds^{2}=d\phi ^{2}+\sin ^{2}\phi \, (d\Theta ^{2}+\sin ^{2}\Theta \, d\Phi ^{2})
                                                                \, \text{.}
\end{equation}

We focus on a particular spherically symmetric harmonic map
(a corotational equivariant map) obtained via the well-known
hedgehog ansatz:
\begin{equation}
\label{hedgehog}
\phi (x^{\mu })=\phi (u,r),\quad \Theta (x^{\mu })=\theta
                                                                \, \text{,}
        \quad
\Phi (x^{\mu })=\varphi
                                                                \, \text{.}
\end{equation}
With this ansatz two of the three coupled fields are determined and only
one field $ \phi (u,r) $ enters the equations.
Regularity at the origin forces the $\sigma$--field $\phi$ to vanish
at $r=0$, so the origin is always mapped to one of the poles of $S^{3}$,
defined by the choice of coordinates (\ref{targetMf-metric}).
As $ \phi  $ represents the ``areal coordinate''
of the polar coordinate system (\ref{targetMf-metric}) on the target manifold,
its regularity behavior near the origin is the same as that of the
areal coordinate $r$:
\begin{equation}
\phi (u_0,r)=O(r)
                                                                \, \text{.}
\end{equation}
The matter field equations
are then reduced to the single nonlinear wave equation
\begin{equation}
\label{eqn-phi}
\square\phi =\frac{\sin (2\phi )}{r^{2}}
                                                                \, \text{,}
\end{equation}
where $ \square $ is the wave operator
$g^{\mu\nu}\nabla_{\mu}\nabla_{\nu}$:
\begin{equation}
\square = e^{-2\beta}\left(\left(\frac{2V}{r^2} +
\left(\frac{V}{r}\right)^\prime \right)\partial_r
- \frac{2}{r}\partial_u - 2\partial_u\partial_r +
\frac{V}{r}\partial_{rr}\right).
\end{equation}

The nontrivial Einstein equations
split up into the hypersurface equations (the
$\{rr\}$ and $\{ur\} - (V/2r)\{rr\}$ components of
$G_{\mu\nu} = \kappa T_{\mu\nu}$)
\begin{mathletters}
                                                        \label{eqn-hypersurface}
\begin{eqnarray}
\beta'  & = & \frac{\ccbeta }{2}r(\phi' )^{2}
                                                                \, \text{,}
                                                        \label{eqn-beta'}
                                                                        \\
V' & = & e^{2\beta }(1-2\ccbeta \sin (\phi )^{2})
                                                                \, \text{,}
                                                        \label{eqn-V'}
\end{eqnarray}
\end{mathletters}
the subsidiary equation ($r^2 (\{uu\} - (V/r) \{ur\})$)
\begin{eqnarray}
{\dot V} -  2 V {\dot\beta} & = &
                2 \ccbeta \left[
                           \bigl( r {\dot\phi} \bigr)^2
                           - \frac{V}{r}
                             \bigl( r \phi' \bigr)
                             \bigl( r {\dot\phi} \bigr)
                           \right]
                                                                \, \text{,}
                                                \label{eqn-Euur-subsidiary}
\end{eqnarray}
and the redundant equation ($\{\theta \theta\}$)
\begin{eqnarray}
V (r \beta'' - \beta')
                        & + & r \beta' V'
                          + \thalf r V''
                          - 2 r^2 \partial_{ur} \beta
                                                                \nonumber\\
\qquad
        & =  &
                  \ccbeta r \phi'
                  ( - V \phi' + 2 r \partial_u \phi )
                                                                \, \text{.}
                                                \label{eqn-Ethth-redundant}
\end{eqnarray}

The combination of the hypersurface equations~\eqref{eqn-hypersurface}
and the matter field equation~\eqref{eqn-phi} suffices to evolve all
the dynamical fields $\Vr$, $\beta$, and $\phi$.  Assuming these equations
to be satisfied, the redundant equation~\eqref{eqn-Ethth-redundant} then holds
identically, and if the subsidiary equation~\eqref{eqn-Euur-subsidiary}
is satisfied
on some $r = \constant$ surface (this is assured for $r = 0$ by the
regularity conditions there), then it too must hold everywhere.

In view of this, we construct initial data on a $u = \constant$
slice by choosing $\phi$ as free data on the slice, then integrating
the hypersurface equations~\eqref{eqn-hypersurface} to obtain the
metric coefficients $\Vr$ and $\beta$ on the slice.  To evolve this
data to future $u = \constant$ slices, we simultaneously integrate
the hypersurface equations~\eqref{eqn-hypersurface} and the matter
field equation~\eqref{eqn-phi}.  Throughout the initial data construction
and the evolution, we use the subsidiary equation~\eqref{eqn-Euur-subsidiary}
and the redundant equation~\eqref{eqn-Ethth-redundant} solely to check
the accuracy of our numerical computations.
We discuss our numerical treatment of all these equations in
appendix~\ref{app-numerical}.

In our coordinates, the Misner-Sharp mass function
(\citesprefix\cite{Misner-Sharp-1964-Lagrangian-spherical-collapse,
Hayward-1994-mass-functions,Hayward-1996-Misner-Sharp-mass-function})
can be written
directly in terms of the metric,
\begin{equation}
m(u,r) \equiv
m_\MS(u,r) =
        \frac{r}{2}
        \left( 1 - \frac{V}{r} e^{- 2 \beta} \right)
                                                                \, \text{.}
                                                        \label{eqn-m-MS-defn}
\end{equation}
or by using the Einstein equations, rewritten as a radial integral
within a single slice,
\begin{mathletters}
                                                        \label{eqn-m-rho-defn}
\begin{eqnarray}
m(u,r)  & \equiv &
                m_\rho(u,r)
                = \int\limits_0^r \, m'(u,{\tilde r}) \, d{\tilde r},
                                                                        \\
\noalign{\hbox{where}}
m'(u,r) & = &
                \frac{\ccbeta}{2} r^2
                \left(
                \frac{V}{r} e^{-2 \beta} (\phi')^2
                + 2 \frac{\sin^2 \phi}{r^2}
                \right).
\end{eqnarray}
\end{mathletters}

Since our coordinates would be singular on an apparent horizon,
we have designed our numerical evolution scheme to slow down as an
apparent horizon is approached, in such a manner that the evolution
only asymptotes to the apparent horizon
(\cf{}~appendix~\ref{app-numerical/overview}).

In other words, none of our numerically-computed slices ever actually
contain an apparent horizon.  Thus strictly speaking we can never
{\em measure\/} a black hole mass, but only {\em estimate\/} what the
mass {\em will\/} be when (if) a black hole eventually forms.  To do
this, at each numerical time step we compute the Misner-Sharp mass
function $m_\MS$, and look for regions of the numerical grid which
are almost at the critical density for black hole formation, \ie{}
where $2m_\MS / r$ is almost~1.  More precisely, if anywhere in the
grid $2m_\MS / r$ exceeds a specified threshold
\footnote{
         0.995 for all results reported here.
         }
, then we estimate that a black hole will form, with a final mass $m_\BH$
given by the mass function $m_\MS$ at the outermost such grid point.
In general this mass estimate changes during the evolution; we
use the last value before a numerical evolution terminates as our
overall estimate for the black hole mass.

It is also of interest to compute the total mass $m_\total$ within the
outer grid boundary.
\footnote{
         To be precise, we use $m_\rho$ at the outermost
         grid point, not $m_\MS$, since $m_\MS$ is numerically
         somewhat ill-conditioned in the outer part of the grid,
         whereas $m_\rho$ is well-conditioned everywhere.
         }
{}  This gives an upper bound for our final estimate $m_\BH$.


\section{Results}
\label{sect-results}

For each value of the coupling constant $\ccbeta$,
we consider a 1-parameter family of initial
data $\phi = \phi_p(u_0,r)$, such that (say) for small values of $p$ this
initial data eventually disperses without forming a black hole, while for
large values of $p$ it eventually forms a black hole.  By using a binary
search in $p$, we can find (a numerical approximation to) the critical
value $p = p^*$ which defines the threshold of black hole formation.

We have studied the Einstein--$\sigma$-model system in this manner over
the range of coupling constants $0.18 \leq \ccbeta \leq 100$, using
several different initial-data families.  Here we present results using
the Gaussian-like initial data family
\begin{equation}
\phi(u_0,r)
        = A r^2 \exp\left[-\left(\frac{r-r_0}{\sigma}\right)^2\right]
                                        \label{eqn-Gaussian-initial-data}
\end{equation}
with the ``amplitude'' $A$ as the parameter~$p$ (holding $\sigma$ and
$r_0$ constant for a given critical search),
and also using the ``derivative of 4th-power pseudo-Gaussian'' family
\begin{equation}
\phi(u_0,r)
        = -4 A r^2
          \left(\frac{r-r_0}{\sigma}\right)^3
          \exp\left[-\left(\frac{r-r_0}{\sigma}\right)^4\right]
                                        \label{eqn-d4g-initial-data}
\end{equation}
with the ``width'' $\sigma$ as the parameter~$p$ (holding $A$ and $r_0$
constant for a given critical search).
\footnote{
         For this latter case the relative sign of~$p$
         is reversed with respect to black hole formation,
         \ie{} for large~$p$ the initial data eventually
         disperses, while for small~$p$ it eventually
         forms a black hole.
         }
{}  All the results reported here used a ``position'' $r_0 = 5$ and
an initial-slice outer boundary of $r_\outer = 30$.
Table~\ref{tab-initial-data} shows some near-critical initial data
parameters.

We have also carried out a number of convergence tests of our numerical
scheme, both for single evolutions and for entire critical searches.
We discuss these in appendix~\ref{app-convergence}.


\subsection{DSS Echoing}
\label{sect-results/echoing}

Discrete self-similarity is defined by the existence of a discrete
diffeomorphism $\Phi_\Delta$ such that
for some fixed $\Delta \in \mathbb{R}$,
\begin{equation}
\left(\Phi_\Delta^*\right)^n g = e^{2 n \Delta} g
        \qquad
                \forall n \in \mathbb{N}
                                                                \, \text{.}
                                                        \label{eqn-DSS-defn}
\end{equation}
In adapted coordinates $\tau = -\ln \frac{u^* - u}{u^*}$ and
$\rho = \frac{r}{u^*-u}$, where $u^*$ is a real number which denotes the
accumulation time of DSS, we have
\begin{equation}
Z(\tau +n\Delta,\rho) = Z(\tau, \rho)
        \qquad
                \forall n \in \mathbb{N}
                                                                \, \text{,}
\end{equation}
where Z denotes $\beta$, $\Vr$, $\phi$, or any combination of these,
\eg{} $2m/r$.
In addition the $\sigma$-field $\phi$ satisfies the stronger condition
\begin{equation}
\left( \Phi_{\Delta/2}^* \right)^n \phi = (-1)^n \phi
                                                                \, \text{,}
                                                        \label{eqn-DSS-Delta/2}
\end{equation}
so that fields even in $\phi$ (\eg{} $\beta$, $\Vr$, and quantities
constructed from them) are actually periodic in $\tau$ with period
$\Delta/2$.  As a DSS diagnostic, we typically look for ($\Delta/2$)
periodicity in the black hole formation diagnostic $\max\,\, 2m/r$,
where the maximum is taken over $r$ within each $u = \constant$ slice.

\begin{figure}[tb]
\begin{center}
\begin{picture}(160,70)
\begin{psfrags}
\put(12,10){
           \begin{picture}(60,60)
           \psfrag{ccbeta = 0.5}{\hskip-0.5em{}Part~(a)}
           \psfrag{-ln(u* - u)}
   {\hskip-1.6em\smash{\lower1.0ex\hbox{$- \ln(u^* - u)$}}}
           \psfrag{max 2m/r}{$\max\,\, \frac{2m}{r}$}
           \put(0.0,0.0){\hbox{\epsfxsize=65mm\epsfbox{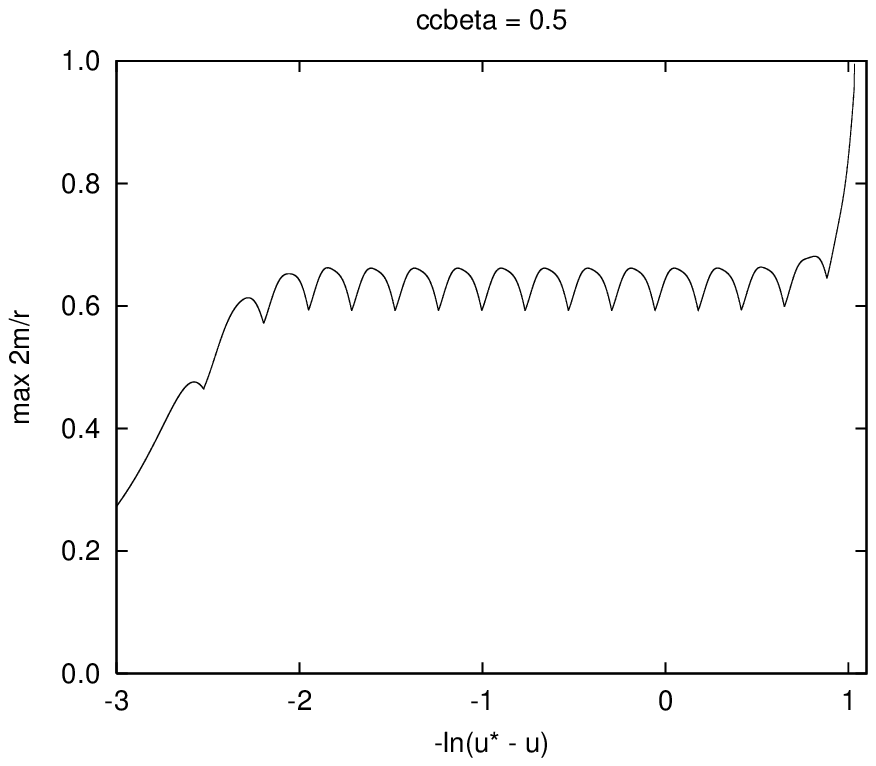}}}
           \end{picture}
           }
\put(97,10){
           \begin{picture}(60,60)
           \psfrag{ccbeta = 0.18}{\hskip-0.5em{}Part~(b)}
           \psfrag{-ln(u* - u)}
   {\hskip-1.6em\smash{\lower1.0ex\hbox{$- \ln(u^* - u)$}}}
           \psfrag{max 2m/r}{$\max\,\, \frac{2m}{r}$}
           \put(0.0,0.0){\hbox{\epsfxsize=65mm\epsfbox{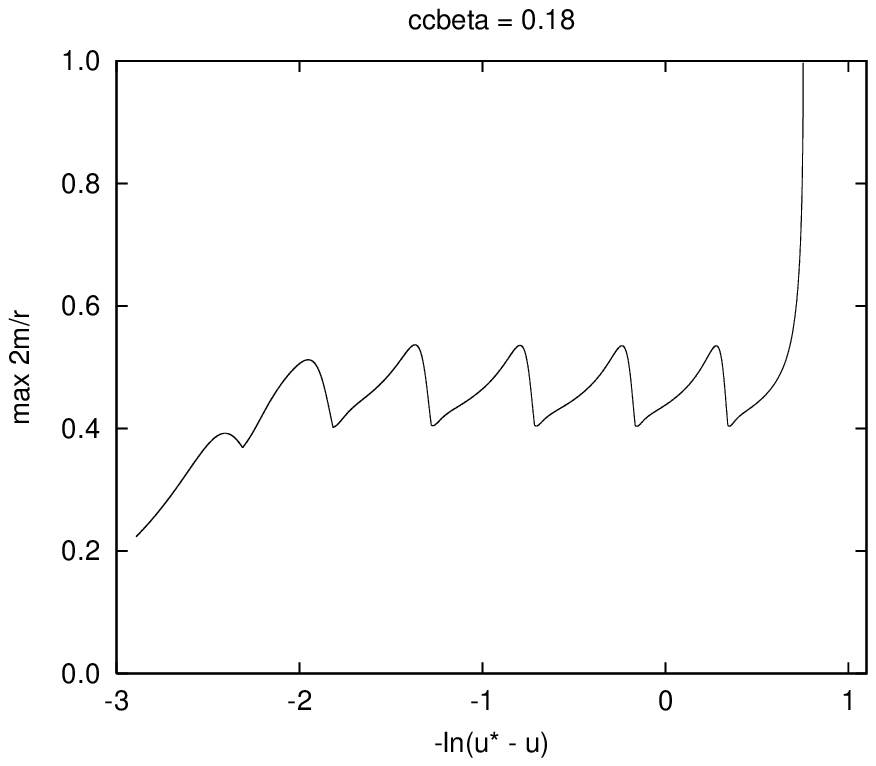}}}
           \end{picture}
           }
\end{psfrags}
\end{picture}
\end{center}
\caption[Echoing of $\max \,\, 2m/r$ in Near-Critical Evolutions]
        {
        This figure shows DSS echoing behavior
        in the black hole formation diagnostic \,$\max\,\, 2m/r$\,
        in near-critical (in this case slightly supercritical)
        evolutions for coupling constants $\ccbeta = 0.5$~(part~(a))
        and $0.18$~(part~(b)).  Notice the much longer period
        $\Delta/2$ of the echoes at $\ccbeta = 0.18$.
        Although it's not apparent to the eye at the scale of
        this figure, the $0.18$ echoes aren't {\em exactly\/}
        identical: they vary in period and shape by \hbox{5--10\%}.
        }
\label{fig-echoing}
\end{figure}
\begin{figure}
\begin{center}
\begin{psfrags}
\epsfxsize=2.2in\leavevmode\rotatebox{-90}{\epsfbox{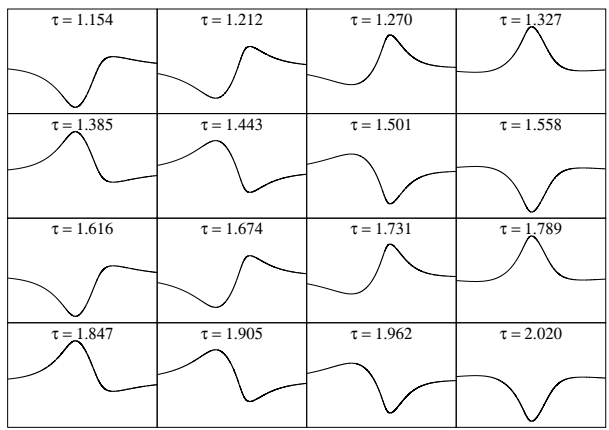}}
\end{psfrags}
\end{center}
\caption{Snapshots of a near-critical evolution of the SU(2)
  $\sigma$-field $\phi$ as a function of $\ln r$ for $\ccbeta = 1.0$.
  The frames are evenly spaced in $\tau = -\ln \frac{u^*-u}{u^*}$
  (this increases towards the accumulation time $u^*$).
  $\tau$ increases to the right within each row, then downwards between
  successive rows.
  Observe that $\phi$ is the same in frames in the same column
  but 2 rows apart; this indicates that $\phi$ is periodic in $\tau$
  with period $\Delta = 0.46$.
  Also notice that $\phi$ is negated between frames in the same column
  of adjacent rows, \ie{} it satisfies the half-period self-similarity
  condition~\protect\eqref{eqn-DSS-Delta/2}.
       }
\label{fig-phi-multi-DSS}
\end{figure}

We have clear evidence for the existence of a type II critical collapse
with a DSS critical solution.  Figure~\ref{fig-echoing} shows examples
of this for two values of the coupling constant.  Since
\,$\max\,\, 2m/r$ periodicity is only a neccesary condition for DSS, we
have also explicitly verified that the matter field $\phi$ at selected
times~$u$ coincides with its image under the DSS diffeomorphism~$\Phi$;
figure~\ref{fig-phi-multi-DSS} shows an example of this.

\begin{figure}[tb]
\begin{center}
\begin{picture}(75,60)
\put(0,0){\epsfxsize=75mm\epsfbox{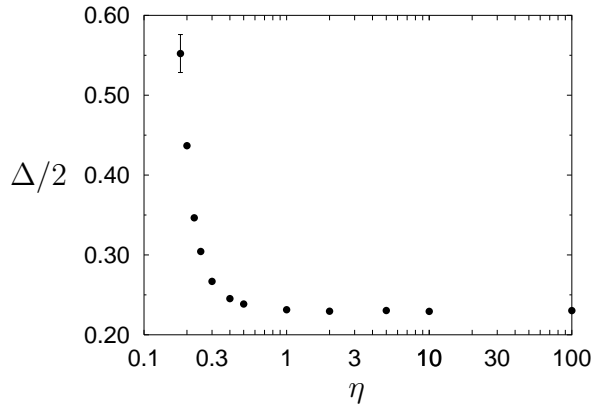}}
\put(40.0,-2.0){$\ccbeta$}
\put(-5.0,26.2){$\Delta/2$}
\end{picture}
\end{center}
\caption[Variation of DSS Self-Similarity Period $\Delta/2$
         with Coupling Constant $\ccbeta$]
        {
        This figure shows the variation of the near-critical
        $\max\,\, 2m/r$ echoing period $\Delta/2$ with the
        coupling constant $\ccbeta$, for the Gaussian-like
	initial data family given in table~\protect\ref{tab-initial-data}.
	Notice the rapid rise in $\Delta/2$ at small $\ccbeta$.
	The error bar for the $\ccbeta=0.18$ point is estimated
	from the dispersion in $\Delta/2$ values when fitting
	different subsets of echoes in
	figure~\hbox{\protect\ref{fig-echoing}(b)}; for
        larger values of $\ccbeta$ this dispersion is negligible.
        }
\label{fig-Delta/2-of-ccbeta}
\end{figure}

We find that the self-similarity echoing period $\Delta/2$ varies
strongly with the coupling constant $\ccbeta$.
Table~\ref{tab-initial-data} gives some numerical data showing this,
and figure~\ref{fig-Delta/2-of-ccbeta} shows this same data graphically.
At large $\ccbeta$, $\Delta/2$ asymptotes to $0.2300 \pm 0.0003$.
As $\ccbeta$ decreases towards the lower limit of
the data in table~\ref{tab-initial-data},
$\ccbeta = 0.18$, $\Delta/2$ rises sharply.
At the very smallest coupling constant $\ccbeta \,{=}\, 0.18$, but not
at $\ccbeta \,{=}\, 0.20$ or any larger value, the critical solution
shows small deviations from exact DSS: the periods and shapes of the
individual $\max\,\, 2m/r$ oscillations deviate by \hbox{5--10\%} from
the best-fitting DSS prediction.
The physical significance of this is not yet clear.


\subsection{Scaling and Universality}
\label{sect-results/scaling}

In the presence of DSS, the black hole mass $m_\BH$ of
slightly-supercritical evolutions shows a universal scaling law
(\citesprefix\cite{Koike-Hara-Adachi-1995-scaling-in-critical-collapse,
Koike-Hara-Adachi-1996-scaling-in-critical-collapse,
Gundlach-1996-scaling-in-critical-collapse,
Hod-Piran-1997-fine-structure-of-mass-scaling-law})
\begin{equation}
\ln m_\BH = \gamma \ln(p-p^*)
            + \Psi \bigl( \ln(p-p^*) \bigr)
            + \constant
                                                        \label{eqn-mass-scaling}
\end{equation}
where $\gamma$ sets the overall slope of the scaling law, and the
function $\Psi$ is periodic with period $\thalf \Delta/\gamma$
in $\ln(p-p^*)$.  For slightly-subcritical evolutions, the maximum
(taken over $u$ within each evolution) of the 4-Ricci scalar evaluated
at the origin, $R_\max$, also shows a similar scaling law,
but with slope $-2\gamma$
(\citeprefix\cite{Garfinkle-Duncan-1998-critical-curvature-scaling}).
This is also true for supercritical evolutions, with $R_\max$ now
defined by taking the maximum in $u$ within each evolution only
until a (null) slice reaches the apparent horizon.
(Our actual evolutions terminate slightly before the apparent horizon,
but $R_\max$ doesn't change significantly in this interval.)

\begin{figure}
\begin{center}
\begin{psfrags}
\psfrag{ln m_total}[][c]{$\:\scriptstyle{\ln m_\total}$}
\psfrag{ln m_BH}[][c]{$@,@,\scriptstyle{\ln m_{\BH}}$}
\psfrag{ln R\_max2}[][c]{$@,@,@,\negthickspace\scriptstyle{\ln R_{\max}}$}
\psfrag{ln R\_max}[][c]{$\ln R_{\max}$}
\psfrag{ln m}[][c]{$\ln m$}
\psfrag{ln(p-p*)}[][c]{$\ln (p-p^*)$}
\epsfxsize=2.2in\leavevmode\rotatebox{-90}{\epsfbox{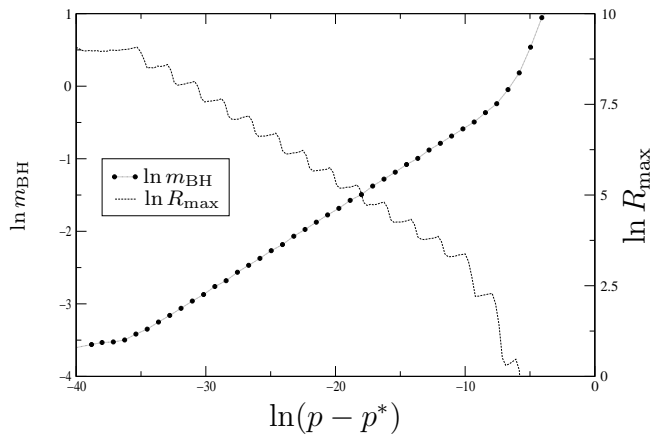}}
\end{psfrags}
\end{center}
\caption{Supercritical scaling of the black hole mass $m_\BH$,
         and of $R_{\max}$, the maximum (over retarded
         time $u$ within a single evolution) of the scalar curvature at the
         origin. The slopes of the masses and $R_{\max}$
         are $+\gamma$ and $-2\gamma$ respectively.
         The scaling fine-structure is clearly visible for $R_{\max}$.
         Its period is found to be $2.099$ which
         is very close to the value $\thalf\Delta/\gamma = 2.097$
         predicted by
         perturbation theory and computed from known values of the critical
         exponents.
         This series of evolutions was done for $\ccbeta = 0.5$ using a
         resolution of 2000 gridpoints.
        }
\label{fig-super-scaling}
\end{figure}

\begin{figure}
\begin{center}
\begin{psfrags}
\psfrag{m_BH}[][c]{$\scriptstyle{m_{\BH}}$}
\psfrag{m_outer}[][c]{$\scriptstyle{m_\total}$}
\psfrag{ln(m) - <ln(m)>}[][c]{$\ln m_\BH - \ln m_{\fit}$}
\psfrag{ln(p-p*)}[][c]{$\ln (p-p^*)$}
\epsfxsize=2.2in\leavevmode\rotatebox{-90}{\epsfbox{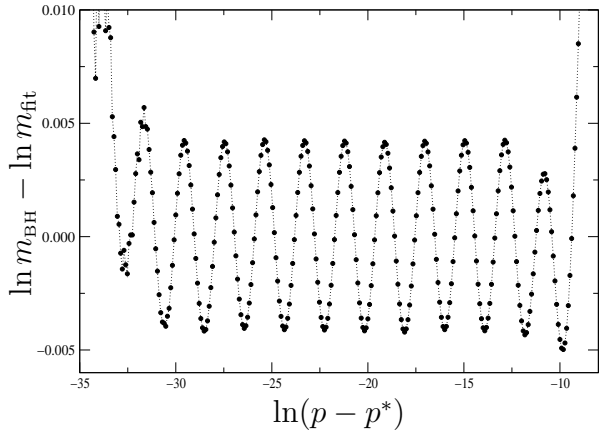}}
\end{psfrags}
\end{center}
\caption{
        This figure shows the fine-scale structure in $m_\BH$
        (shown in figure~\protect\ref{fig-super-scaling}) after subtracting
        a linear fit.  For these evolutions we disabled the
        ``$2m_\MS/r > 0.995$ detected for $N$ time steps'' stopping criterion
        in our code (\cf{}~section~\protect\ref{app-numerical/diagnostics}),
        running each evolution until $\Delta u < 10^{-15}$;
        in this case our final slices' outer grid boundaries almost touched
        the apparent horizon, so $m_\BH$ and $m_\total$ were essentially
        identical (within $\ltsim 10^{-10}$ of each other).
        }
\label{fig-masses-fine-structure}
\end{figure}

We have investigated these scaling laws using a sequence of supercritical
evolutions
with varying $\ln (p-p^*)$.  We extract $\gamma$ by least-squares
fitting $\ln m_\BH$ as a linear function of $\ln(p-p^*)$; after
subtracting this fit from $\ln m_\BH$, we are left with the periodic
fine structure.
Figure~\ref{fig-super-scaling} shows a typical supercritical scaling law
and figure~\ref{fig-masses-fine-structure} its fine structure.

Since the numerical resolution of our code is limited by the use of
IEEE double precision floating point numbers, we expect the errors
to blow up for $p-p^* \ltsim 10^{-16}$ which corresponds to
$\ln(p-p^*) \ltsim 35$.  This can be seen in figures~\ref{fig-super-scaling}
and~\ref{fig-masses-fine-structure}, and also in
figure~\hbox{\ref{fig-pstar-convergence}(b)} (discussed in the next section).
For $p-p^* \gtsim -10$ deviations from the scaling laws are also
apparent, demarcating the range of validity of linear perturbation
theory.

We find that the mass scaling exponent $\gamma$ varies by at most
$5\%$ over the range of $\ccbeta$ we have studied, asymptoting to
$\gamma = 0.1185 \pm 0.0005$ at large $\ccbeta$.  (The error is
estimated from the dispersion in $\gamma$ values between fits to
critical searches with different initial-data families and/or finite
difference grid resolutions.)

\begin{figure}
\begin{center}
\begin{psfrags}
\psfrag{ccbeta}[][c]{$\ccbeta$}
\psfrag{delta/2gamma}[b][c]{$\thalf\Delta/\gamma$}
\epsfxsize=2.2in\leavevmode\rotatebox{-90}{\epsfbox{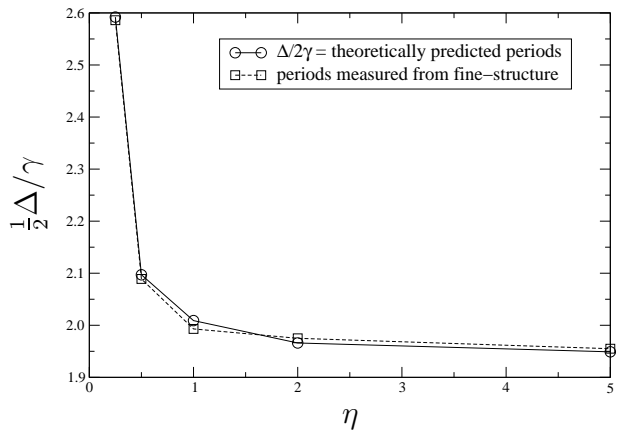}}
\end{psfrags}
\end{center}
\caption{This figure compares the quantity $\thalf\Delta/\gamma$,
         as computed from
         the echoing in $\max~2m/r$ and the mass-scaling law, to the period of
         the oscillations present in the fine-structure of the mass-scaling law,
         which is predicted by perturbation theory to be $\thalf\Delta/\gamma$.
         This has been carried out for coupling constants ranging from $0.25$
         up to $5$. All evolutions were done with $2000$ gridpoints radial
         resolution.
        }
\label{fig-check-fine-structure}
\end{figure}

The periodicity present in the fine structure of the scaling law
(figures~\ref{fig-super-scaling} and~\ref{fig-masses-fine-structure})
can be measured directly.  Figure~\ref{fig-check-fine-structure}
shows a comparison of the measured periods with the perturbation-theory
prediction $\thalf\Delta/\gamma$ in $\ln(p-p^*)$.  The agreement
is excellent.

Comparing results for different one-parameter families of initial
data, we find that the critical behavior is universal at all
coupling constants $\ccbeta$:  The critical exponent $\gamma$ and
the echoing period $\Delta/2$ are the same for all critical searches
at a given coupling constant, regardless of which initial data family
is used.  For example, table~\ref{tab-initial-data} shows that
$\gamma$ and $\Delta/2$ are the same (to within numerical errors)
for even the very different initial data
families~\eqref{eqn-Gaussian-initial-data}
and~\eqref{eqn-d4g-initial-data}.


\section{Conclusions}
\label{sect-conclusions}

In this paper we have presented a detailed numerical analysis of SU(2)
$\sigma$-models coupled to gravity in spherical symmetry for a wide range
of the coupling constant $\ccbeta$.  For $0.18 \leq \ccbeta \leq 100$
we have evidence of universal critical type~II collapse behavior.
The critical solution is DSS.
We have observed both the typical mass scaling at the threshold of black
hole formation of supercritical initial data, and the corresponding
scaling of the scalar curvature for both sub- and supercritical evolutions.

Our numerical results are based on an outgoing--null-cone formulation
of the Einstein-matter equations, specialized to spherical symmetry
(our numerical methods are discussed in detail in appendix A).
we have carried out thorough convergence tests to ensure the validity
of our results (see appendix~\ref{app-convergence}).
Notably, we have demonstrated second order
uniform-in-$r$ convergence of the error diagnostic $\delta m$ (measuring
finite differencing errors in the Misner-Sharp mass function)
for even very-nearly-critical spacetimes. We have also demonstrated
second order convergence for the initial data's critical parameter $p^*$.
To our knowledge this is the first time the latter has been reported.

In the limit of large couplings our model
corresponds to the $\sigma$-model with 3-dimensional flat
target manifold. This model has already been studied by
Liebling \cite{Liebling-inside-global-monopoles}, where he
considered an additional potential. 
As this potential does not play a role for criticality we should
observe the same critical solution for large couplings.
In fact our results for both the echoing period $\Delta = 0.4604$ and the 
scaling exponent $\gamma = 0.1187$ are in good
agreement with the results reported in
\cite{Liebling-inside-global-monopoles}.

While we observe at most a small variation of the critical exponent
$\gamma$ over the range of coupling constants studied, the period $\Delta$
of the DSS depends strongly on the value of the coupling constant: as
$\ccbeta$ tends to 0.18 from above the period increases by more than
a factor of~2 in the narrow range of $0.18 \leq  \ccbeta \leq 0.3$.
Also, close to the lower limit we observe small deviations from exact
self-similarity.

These observations seem to signal a transition
region around the value of $\ccbeta =  0.18$.  From results of the
work on the $\sigma$-model in flat space
(\citesprefix\cite{Bizon-1999-existence-of-self-similar-sigma-CSS-solutions,
Bizon-Chmaj-Tabor-1999-sigma-3+1-evolution,
Liebling-Hirschmann-Isenberg-1999-sigma-critical})
it is known that there exists a critical (threshold) CSS solution.
In a recent paper, \Bizon{} and Wasserman
(\citeprefix\cite{Bizon-Wasserman-2000-CSS-exists-for-nonzero-beta})
have shown numerically
that this solution persists when gravity is turned on, at least up to
a certain value of the coupling constant.  Whether or not the CSS solution
plays a role at the threshold of black hole formation for small couplings
is under current investigation.


\section*{Acknowledgments}

This work has been supported by
the Austrian Fonds zur F\"{o}rderung der wissenschaftlichen Forschung
(project P12754-PHY), NSF grant PHY 9510895 to the University of
Pittsburgh [S.H.], the Fundacion Federico [M.P.~and P.C.A.],
and G.~Rodgers and J.~Thorn [J.T.].
We thank Piotr \Bizon{} for stimulating discussions, and for
providing us with research results in advance of publication.
S.H.~also thanks Peter H\"ubner for stimulating discussions.
We thank S. L. Liebling for drawing our attention to the fact that
our numerical results for large couplings match those of his work in
\cite{Liebling-inside-global-monopoles}. 


\appendix


\section{Numerical Methods}
\label{app-numerical}


\subsection{Overview}
\label{app-numerical/overview}

We discretize the coupled Einstein-matter equations using second order
finite differencing in $r$ within each $u = \constant$ slice, and in
$u$ along ingoing null geodesics.
Our grid points are generically distributed non-uniformly within each slice:
On the initial slice we place them equidistantly in $r$ between the origin
and some finite maximum radius $r_\outer$, but thereafter they free-fall
in towards the origin along ingoing null geodesics.  We always maintain
a grid point at the origin $r = 0$; when another grid point
reaches the origin we drop the point previously at the origin from the grid.
%

The choice of freely-falling grid points provides some degree of
adaptive grid refinement by the focusing of geodesics towards
regions of strong curvature.  Following Garfinkle
(\citeprefix\cite{Garfinkle-1995-sssf-2+2-self-similarity}), we
also gain additional resolution at late times by explicitly refining
our grid by a factor of two everywhere in the slice, each time we have
lost half of the remaining grid points.
Again following \citeprefix\cite{Garfinkle-1995-sssf-2+2-self-similarity},
for some runs we also manually fine-tune the position of the outermost
grid point on the initial slice ($r_\outer$), so that this grid point
will eventually almost hit the strongest-field region of spacetime.
This greatly improves the effectiveness of the factor-of-two grid
refinements, but this method was not required for the results
presented here.
\footnote{
         By fine-tuning $r_\outer$ in this way, we have also
         observed DSS in the massless scalar (Klein-Gordon)
         field, with up to 5~echoes visible (in the sense
         of figure~\ref{fig-echoing}).  This provides a strong
         additional test of our numerical scheme, since the
         dynamic range of the DSS is much larger in the
         Klein-Gordon case: $\Delta/2 \approx 1.73$ there
         (as defined by~\eqref{eqn-DSS-defn}),
         much larger than the values we find for the
         $\sigma$-field.
         }

By moving our grid points along null geodesics, the physical domain
of dependence is automatically contained in the numerical domain of
dependence, so our time step is not restricted by the usual
Courant-Friedrichs-Lewy (CFL) stability limit
(\citesprefix\cite{Courant-Friedrichs-Lewy-1928,Courant-Friedrichs-Lewy-1967}).
However, in order to control time resolution we require (following
\citesprefix\cite{Goldwirth-Piran-1987-sssf-2+2,
Goldwirth-Ori-Piran-1989-in-Frontiers}),
that
\begin{equation}
\parVr \, \Delta u \leq C \, \Delta r
                                                \label{eqn-Delta-u-limit}
\end{equation}
everywhere in the grid, where $C$ is a constant which we typically
take to be on the order of unity.  The time step $\Delta u$ is thus
limited such that grid points fall inwards by no more than $C/2$
grid point spacings per time step.  Most
of our results reported here were obtained with $C = 1.5$.
[Note that for a null-cone evolution similar to ours, but with grid
points at constant~$r$
(\citeprefix\cite{Gomez-Winicour-1992-sssf-2+2-asymptotics}),
there {\em is\/} a CFL stability limit, which is in fact
just~\eqref{eqn-Delta-u-limit} with $C = 2$.]

For $r \, \Theta_+$ sufficiently small,
a large value of $V/r$ decreases the time step $\Delta u$ as follows:
{}From~\eqref{Theta-pm} it is clear that for small $r \, \Theta_+$
the function $\beta$ -- which is monotonically increasing with~$r$  --
becomes large (it blows up at an apparent horizon).  Furthermore,
by~\eqref{eqn-m-MS-defn} we get $V/r = e^{2 \beta} \, (1 - 2 m/r)$.
Outside of the outermost local maximum of $2 m/r$ both $e^{2 \beta}$ and
$1 - 2m/r$ are monotonically increasing with~$r$, and thus so is $V/r$.
If the outer boundary of the grid is taken sufficiently far out, this
is therefore the location of the maximum of $V/r$, and thus of the most
stringent slowdown condition. If $\Delta u < 10^{-15}$ (\ie{} close
to machine precision) the evolution is terminated.

For the remainder of this appendix,
we adopt the usual notation where superscripts denote ``temporal''
($u$) levels.
Figure~\ref{fig-grid-points} shows the typical organization of our grid.
All discretizations in time~($u$) and space~($r$) use nonuniform grid
spacings to allow for the free fall of the gridpoints and the adaptive time
stepping~\eqref{eqn-Delta-u-limit}.
Our numerical scheme uses the geometry fields $\beta$, $\Vr$ and
$\parVrp$, and the rescaled matter field $\psi \equiv r \phi$.
%

Assuming that these fields are known at all grid points on the
$u = u^k$ and $u = u^{k-1}$ slices, we determine the fields on the
$u = u^{k+1}$ slice as follows:
\begin{itemize}
\item   For the innermost 3~non-origin grid points in the $u = u^{k+1}$
        slice, we use a Taylor series expansion as described
        in section~\ref{app-numerical/Taylor-series}.
\item   We then sweep outwards over the remaining spatial grids
        of the $u = u^{k+1}$ slice as discussed in
        section~\ref{app-numerical/evolution-schemes}.
\end{itemize}




\subsection{Taylor Expansions near the Symmetry Axis}
\label{app-numerical/Taylor-series}

The coupled Einstein-matter equations and regularity determine the
generic         
behavior of $\psi$ near the origin as
\begin{equation}
\psi(u + \Delta u, r)
        = c_1 r^2 + c_2 r^3 + c_2 r^2 \, \Delta u
          + O \bigl( \Delta u^4 + r^4 \bigr)
                                                \label{eqn-psi-Taylor-series}
                                                                \, \text{.}
\end{equation}
Substitution of this series expansion into the hypersurface
equations~\eqref{eqn-hypersurface} yields corresponding series
expansions for the geometry fields $\beta$ and $\Vr$.

\begin{figure}
\begin{center}
\begin{picture}(100,80)
\put(20,10){\epsfbox{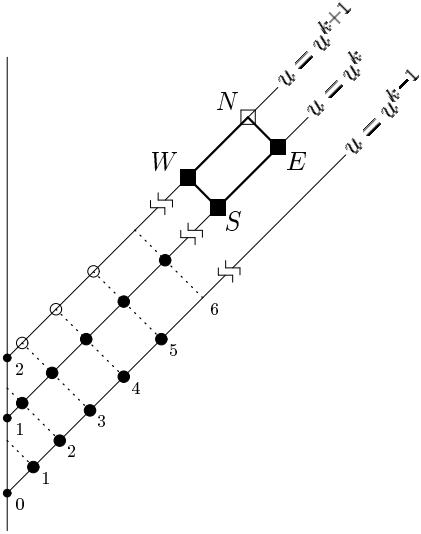}}
\end{picture}
\end{center}
\caption[The Finite Differencing Grid]
        {
        This figure shows our finite differencing grid.
        Individual grid points are labelled with integers~0 to~6,
        and their ingoing-null-geodesic trajectories are shown
        as dotted lines.
        Grid points at the origin are marked with small points.
        Grid points used in the least-squares fitting procedure
        (\cf{}~section~\protect\ref{app-numerical/Taylor-series})
        are marked with large solid circles, while the grid points
        where the field variables are calculated from the Taylor
        series are marked with large open circles.  Grid points
        where $\psi$ is already known in the diamond-integral
        scheme (\cf{}~section~\protect\ref{app-numerical/evolution-schemes})
        are marked with solid squares, while the grid point
        where $\psi$ is computed in this scheme is marked with
        an open square.
        }
\label{fig-grid-points}
\end{figure}

To determine the geometry and matter fields near the origin on
the $u = u^{k+1}$ slice,
we first least-squares fit the functional form~\eqref{eqn-psi-Taylor-series}
to the numerically
computed $\psi$ values at the 5~innermost non-origin points of
the $u = u^k$ and $u = u^{k-1}$ time levels (these points are marked
by large solid circles in figure~\ref{fig-grid-points}).
This determines the coefficients $c_1$ and $c_2$.

For each of the 3~innermost non-origin grid points on the $u = u^{k+1}$
slice (these points are marked by open circles in figure~\ref{fig-grid-points}),
we first integrate the ingoing null geodesic equation~\eqref{eqn-geodesic}
from $u = u^k$ to $u = u^{k+1}$, as described below.
Then, using the coefficients $c_1$ and $c_2$, we determine
$\psi$ at this grid point from the series
expansion~\eqref{eqn-psi-Taylor-series}. Finally, we compute $\beta$,
$\Vr$, and $\parVrp$ from their corresponding series expansions.


\subsection{Integration Schemes}
\label{app-numerical/evolution-schemes}


In order to integrate out from the Taylor series region to the
outer boundary, our general strategy at each grid point is as follows:
\begin{itemize}
\item   We first determine the grid point's $r$ coordinate
        on the $u = u^{k+1}$ slice by integrating the ingoing
        null geodesic equation~\eqref{eqn-geodesic} from
        $u = u^k$ to $u = u^{k+1}$.
\item   We then determine $\psi$ at this grid point using
        a ``diamond integral'' scheme of \Gomez{} and Winicour
(\citesprefix\cite{Gomez-Winicour-1992-sssf-2+2-asymptotics,
Gomez-Winicour-1992-sssf-2+2-numerical-methods,
Gomez-Winicour-1992-in-dInverno}).
\item   We compute the geometry fields by integrating
        the hypersurface equations one grid point
        outwards on the $u = u^{k+1}$ slice.
\end{itemize}

For the hypersurface equations~\eqref{eqn-hypersurface} and the geodesic
equation~\eqref{eqn-geodesic} we use a second order iterated Runge-Kutta
scheme (adapted from section~5.2.1, equation~(5.6), of
\citeprefix\cite{Hyman-1989-MOL-in-Buchler}).
For a generic ODE system $d\y/dx = \f(x,\y)$ the scheme is as follows:
\begin{subequations}
\begin{eqnarray}
\y^{k+1}_\pred   & = &   \y^k + \Delta x \, \f(x^k,\y^k)    \\
\y^{k+1}         & = &   \y^k + \thalf \, \Delta x
                                \left(
                                \f(x^k,\y^k) + \f(x^{k+1},\y^{k+1}_\pred)
                                \right)
                                                \label{eqn-IRK2-corrector}
\end{eqnarray}
                                                                \label{eqn-IRK2}
\end{subequations}
While this allows straightforward integration of the
hypersurface equations~\eqref{eqn-hypersurface},
the geodesic equation~\eqref{eqn-geodesic} needs special care:
The corrector~\eqref{eqn-IRK2-corrector} requires evaluating
the right-hand-side function~$\f$ at the $x^{k+1}$ time level.  For
the geodesic equation this requires knowing the field $\Vr$ on the
$u = u^{k+1}$ slice, which is not yet computed at the time the geodesic
integration is done. We thus linearly extrapolate
the needed $\Vr$ value from $\Vr$ and $\parVrp$ values one spatial grid
point inwards on the same ($u = u^{n+1}$) slice.

The matter field equation is integrated using a ``diamond integral'' scheme
of \Gomez{} and Winicour
(\citesprefix\cite{Gomez-Winicour-1992-sssf-2+2-asymptotics,
Gomez-Winicour-1992-sssf-2+2-numerical-methods,
Gomez-Winicour-1992-in-dInverno}).
The basic idea is to integrate the nonlinear wave
equation~\eqref{eqn-phi} over the null parallelogram $\Sigma$ spanned
by the 4~grid points $N$, $S$, $W$, and $E$ in figure~\ref{fig-grid-points}.
This allows the nonlinear wave equation~\eqref{eqn-phi} to
be written as
\begin{eqnarray}
\psi(N)
        & = &
                \psi(W) + \psi(E) - \psi(S)
                                                                \nonumber \\
        &   &
                \quad
                - \half
                  \int\limits_\Sigma
                  \left(
                    \left( \frac{V}{r}\right)' \, \frac{\Psi}{r}
                    +
                    e^{2\beta} \,\sin\left(2 \frac{\psi}{r}\right)
                  \right) \, du \, dr
                                                \label{eqn-diamond-integral}
                                                                \, \text{.}
\end{eqnarray}
We evaluate the integral numerically by approximating the integrand
as constant over the null parallelogram $\Sigma$, with a value which
is the average of its values at the grid points $W$ and $E$.  This
gives second order overall accuracy for $\psi$.


\subsection{Diagnostics}
\label{app-numerical/diagnostics}

Within a single evolution, we use several diagnostics to assess
the accuracy of our numerical computations.  We numerically check
the satisfaction of the subsidiary and redundant Einstein
equations~\eqref{eqn-Euur-subsidiary} and ~\eqref{eqn-Ethth-redundant}.
We also compare the two ``different'' forms of the mass function
$m_\MS$ and $m_\rho$:  These are in fact identical by virtue of the
Einstein equations, but they are computed in very different ways
(via~\eqref{eqn-m-MS-defn} and~\eqref{eqn-m-rho-defn} respectively),
and numerically they will generally
differ by a small amount due to finite differencing
errors.  This difference is a useful diagnostic of the code's accuracy.
To this end, we define
\begin{equation}
\delta m(u,r) = \frac{m_\MS - m_\rho}{m_{\total,\,\init}}
                                                \label{eqn-delta-m-defn}
\end{equation}
where $m_{\total,\,\init} \equiv m_\MS(u{=}0, r_\max)$ is the total mass of
our initial slice.  $\delta m$ is then a dimensionless diagnostic of how
well our field variables approximate the Einstein equations; we must
have $|\delta m| \ll 1$ everywhere in the grid at all times
for our results to be trustworthy.


\section{Convergence Tests}
\label{app-convergence}

We use convergence tests of the type popularized by Choptuik
(\citesprefix\cite{Choptuik-PhD,Choptuik-1991-consistency,
Choptuik-Goldwirth-Piran-1992-sssf-cmp-3+1-vs-2+2}) both to better
understand the performance of our numerical algorithms, and to
quantitatively assess the accuracy of our numerical results.
In particular, it is only through such convergence tests that we can
be confident our conclusions reflect properties of the continuum
Einstein-matter equations, rather than numerical artifacts.

\begin{figure}
\begin{center}
\begin{psfrags}
\begin{picture}(160,80)
\psfrag{linear scale}{\hskip-0.7mm{}Part~(a)}
\psfrag{p - p*[inf]}{\hskip-3.5mm{}$p - p^*[\infty]$}
\psfrag{p}{$p$}
\psfrag{m_BH}{$m_\BH$}
\put(0,0){\epsfxsize=75mm\epsfbox{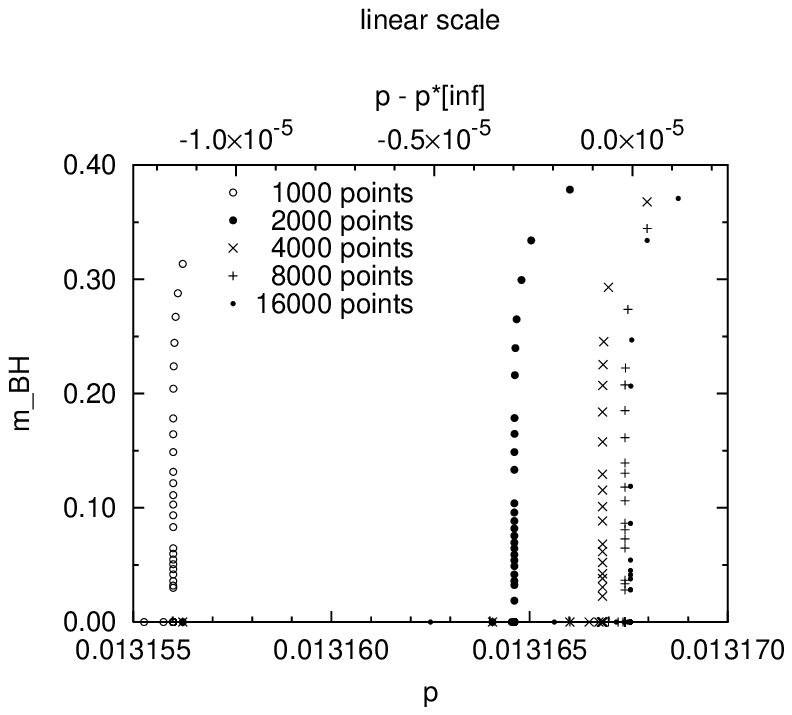}}
\psfrag{log scale}{\hskip-2.3mm{}Part~(b)}
\psfrag{p - p*[N]}{\hskip-3.5mm{}$p - p^*[N]$}
\psfrag{ln (p - p*[N])}{\hskip-3.5mm{}$\ln (p - p^*[N])$}
\psfrag{ln m_BH}{$\ln m_\BH$}
\put(80,0){\epsfxsize=75mm\epsfbox{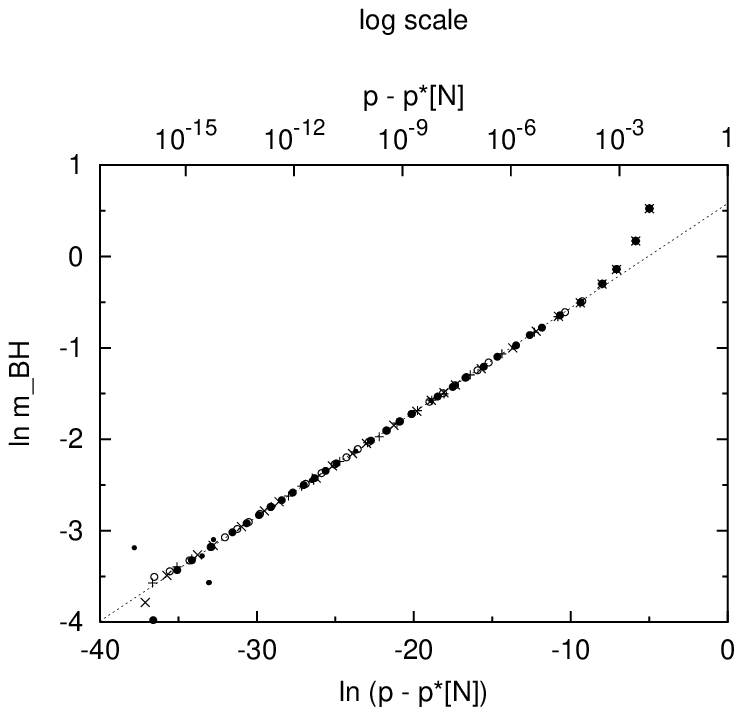}}
\end{picture}
\end{psfrags}
\end{center}
\caption[Convergence of $p^*$ with Grid Resolution]
        {
        This figure shows the convergence of the supercritical
        mass-scaling law with increasing finite difference grid
        resolution, for $\ccbeta = 0.5$ evolutions.
        Part~(a) shows the mass scaling behavior for 5~different
        grid resolutions, plotted on {\em linear\/} scales in both
        $p$ (here the ``amplitude'' $A$) and $m_\BH$.
        Notice how the main effect of changes in the grid resolution
        is to simply shift the entire critical curve to a slightly
        different $p^*$.  The actual $p^*$ values are given in
        table~\protect\ref{tab-pstar-convergence}, and all these evolutions
        used the same initial data parameters as given
        in~table~\protect\ref{tab-initial-data}.
        Part~(b) shows the same data plotted on logarithmic scales,
        with each resolution's $p$ values being taken relative to
        that resolution's own $p^*$ value.  Notice that all the
        resolutions satisfy the same scaling law, even down to
        $p - p^*[N]$ far smaller than the resolution shifts
        shown in part~(a); this is discussed further in the text.
        }
\label{fig-pstar-convergence}
\end{figure}

As an example of the convergence properties of our computational
scheme, we discuss a series of near-critical $\ccbeta = 0.5$ evolutions.
We begin by considering the effects of varying grid resolutions
(specified by the number of grid points $N$) on
the critical parameter $p^*$.  Figure~\hbox{\ref{fig-pstar-convergence}(a)}
shows these effects for the supercritical mass-scaling law.  Notice
that the dominant effect is to simply shift each entire critical curve
to a slightly different $p^*[N]$.  Table~\ref{tab-pstar-convergence}
shows these $p^*[N]$ values, and their convergence to a continuum
limit (which we denote by $p^*[\infty]$) as the grid resolution
is increased.  Notice that the ratios of the successive differences
$(p^*[2N] - p^*[N]) \big/ (p^*[4N] - p^*[2N])$ are very nearly
equal to~4, \ie{} the $p^*$ values show second order convergence to
$p^*[\infty]$.

Besides shifting the effective $p^*$, what other effects does varying
the grid resolution have on the critical behavior?
Figure~\hbox{\ref{fig-pstar-convergence}(b)} shows the same data as
figure~\hbox{\ref{fig-pstar-convergence}(a)}, but plotted using the usual
logarithmic mass-scaling-law coordinates, and with each grid resolution's
data plotted using that resolution's own $p^*[N]$ value.  It is
clear that the different resolutions all yield the same mass scaling
law.

[In order to get the same mass scaling law at different resolutions,
it is essential here to use each resolution's own $p^*[N]$ value, since
figure~\hbox{\ref{fig-pstar-convergence}(b)} shows the mass scaling
law continuing down to $p - p^*[N]$ values some 10~orders of magnitude
smaller than the typical $p^*[N]$ shifts from one resolution to another.
Equivalently, if we did {\em not\/} use each resolution's own $p^*[N]$
value in figure~\hbox{\ref{fig-pstar-convergence}(b)}, then the mass
scaling law would fail to hold below $p - p^*[N] \sim 10^{-5}$
(the typical $p^*[N]$ shifts seen in
figure~\hbox{\ref{fig-pstar-convergence}(a)}), whereas by using each
resolution's own $p^*[N]$ value, it actually continues down to
$p - p^*[N] \sim 10^{-15}$.]

\begin{figure}
\begin{center}
\begin{psfrags}
\begin{picture}(160,160)
\put(50,80){
           \begin{picture}(0,0)
           \psfrag{max 2m/r for delta m convergence}{\hskip14.0mm{}Part~(a)}
           \psfrag{u}{$u$}
           \psfrag{max 2m/r}{\smash{\lower2mm\hbox{\hskip-3mm{}$\max\,\, 2m/r$}}}
           \put(10.0,40.1)
   {\scriptsize 8000 points, $p = p^*[8000] \!+\! 10^{-12}$}
           \put(10.0,35.5)
   {\scriptsize 16\,000 points, $p = p^*[8000] \!+\! 10^{-12}$}
           \put(10.0,30.9)
   {\scriptsize 16\,000 points, $p = p^*[16\,000] \!+\! 10^{-12}$}
           \put(37.5,8.0){\rotatebox{90}{\footnotesize $u = 13.08$}}
           \put(54.6,8.0){\rotatebox{90}{\footnotesize $u = 18.59$}}
           \put(-12.7,-8.5){\epsfbox{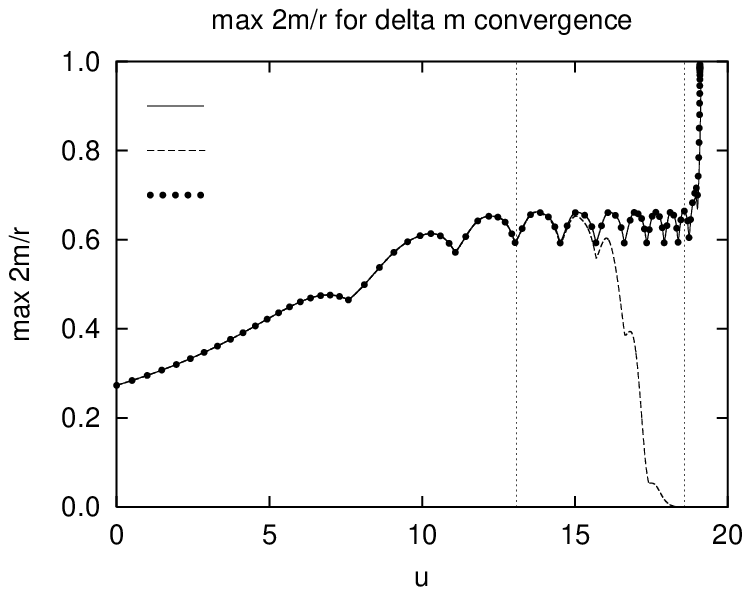}}
           \end{picture}
           }
\put(10,10){
           \begin{picture}(0,0)
           \psfrag{delta m convergence at u = 13.0833}{\hskip14.0mm{}Part~(b)}
           \psfrag{r}{$r$}
           \put(-10.5,22.3){$\delta m$}
           \put(10.0,41.1){\scriptsize $\delta m[8000]$}
           \put(10.0,36.1){\scriptsize $\delta m[16\,000]$}
           \put(10.0,31.1){\scriptsize $4 \times \delta m[16\,000]$}
           \put(2.5,25.0){\fbox{\footnotesize $u = 13.08$}}
           \put(-37.0,-8.5){\epsfbox{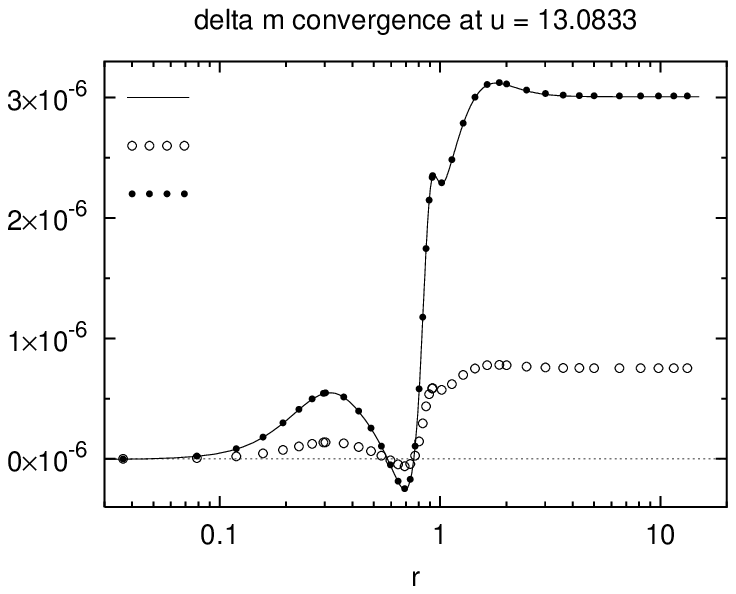}}
           \end{picture}
           }
\put(90,10){
           \begin{picture}(0,0)
           \psfrag{delta m convergence at u = 18.5915}{\hskip14.0mm{}Part~(c)}
           \psfrag{r}{$r$}
           \put(-10.5,28.6){$\delta m$}
           \put(2.5,36.9){\fbox{\footnotesize $u = 18.59$}}
           \put(-40.4,-8.5){\epsfbox{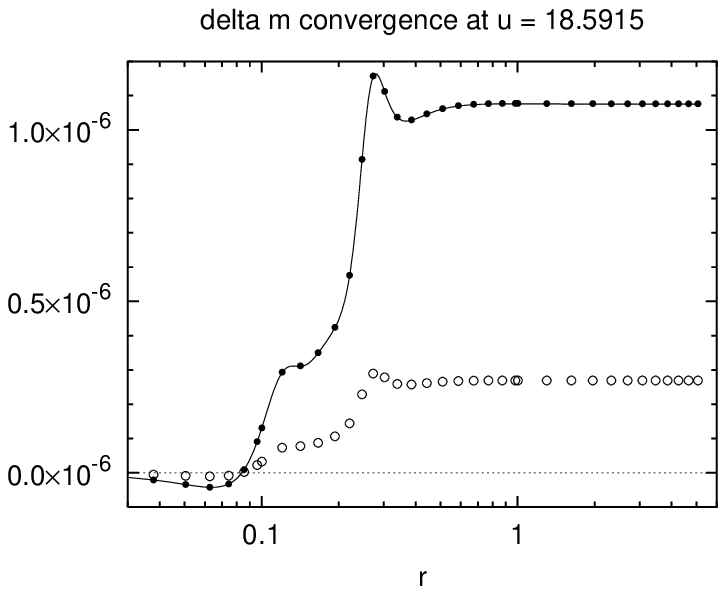}}
           \end{picture}
           }
\end{picture}
\end{psfrags}
\end{center}
\caption[Convergence of $\delta m$ with Grid Resolution]
        {
        This figure shows the convergence of $\delta m$ to zero
        with increasing grid resolution, for near-critical
        $\ccbeta = 0.5$ evolutions.
        Part~(a) shows the time development of $\max\,\, 2m/r$
        for each of 3~evolutions described in the text.
        Part~(b) shows the convergence of $\delta m$ between
        evolutions~(1) and~(2), at a relatively early time.
        Part~(c) shows the convergence of $\delta m$ between
        evolutions~(1) and~(3), at a relatively late time.
        (Note that in all cases, the marked points are spaced
        for ease of reading, and represent only a small subset
        of the time steps or spatial grid points.)
        }
\label{fig-delta-m-convergence}
\end{figure}

We now consider convergence behavior within a single evolution, or
more precisely between the 3~evolutions whose $\max\,\, 2m/r$ time
developments are shown in figure~\hbox{\ref{fig-delta-m-convergence}(a)}:
\begin{enumerate}
\item[(1)]
        The first evolution uses 8000 grid points, with
        $p = p^*[8000] + 10^{-12}$, so this evolution is just
        slightly supercritical, by about 1~part in $10^{10}$.
        This can be seen in the $\max\,\, 2m/r$ plot:
        $\max\,\, 2m/r$ first oscillates a number of times,
        then eventually rises to~1.
\item[(2)]
        The second evolution uses \hbox{16\,000} grid points, with the same
        $p$ as evolution~(1).  Due to the shift in the effective
        $p^*$ with $N$, this evolution is now subcritical,
        in fact subcritical by a relatively large amount:
        $\max\,\, 2m/r$ oscillates only about half as many times
        as in evolution~(1), then eventually decays to zero.
\item[(3)]
        The third evolution also uses \hbox{16\,000} grid points,
        but this time $p$ is adjusted to compensate for the shift in the
        effective~$p^*$ with~$N$: we take $p = p^*[16\,000] + 10^{-12}$.
        By construction, this evolution is supercritical again,
        by the same amount as evolution~(1); in fact its
        $\max\,\, 2m/r$ plot is almost identical to that of
        evolution~(1).
\end{enumerate}

We use $\delta m$ as a diagnostic of our code's numerical accuracy
for these evolutions.
Figure~\hbox{\ref{fig-delta-m-convergence}(b)} shows the convergence
of $\delta m$ to zero for evolutions~(1) and~(2).  These evolutions
eventually yield very different spacetimes (one forming a black hole,
the other not), but here we consider $u = \constant$ slices at an
early enough time, $u = 13.08$ (shown by the left vertical dashed
line in figure~\hbox{\ref{fig-delta-m-convergence}(a)}) that the
evolutions have not drifted very far apart yet.  
From figure~\hbox{\ref{fig-delta-m-convergence}(b)} it is clear
that $\delta m$ is almost precisely a factor of~4 smaller at the higher
resolution than at the lower one, \ie{} $\delta m$ shows second order
convergence to zero, as expected from the construction of our finite
differencing schemes.  Notice also that this convergence is
{\em uniform\/}, which is a considerably stronger numerical-fidelity
requirement than requiring only pointwise or gridwise-norm convergence.

Now consider a convergence test between evolutions~(1) and~(3).
Because evolution~(3) adjusts $p$ to compensate for the shift in the
effective $p^*$ with~$N$, these two evolutions have very similar
behavior, so we can consider much later $u = \constant$ slices and
still obtain good convergence.  For example,
figure~\hbox{\ref{fig-delta-m-convergence}(c)} shows the convergence
of $\delta m$ at the relatively late time $u = 18.59$ (shown by the
right vertical dashed line in figure~\hbox{\ref{fig-delta-m-convergence}(a)}).
The convergence is (again) very accurately second order.  In other
words, once we compensate for the shift in the effective $p^*$ with~$N$,
we have excellent -- and uniformly pointwise -- convergence even for
evolutions that are {\em very\/} close to critical ($p - p^*[N]$ here
is about 5~orders of magnitude smaller than the $p^*[N]$ shifts between
the two resolutions), and hence {\em very\/} sensitive to small
perturbations.


\bibliography{jt}

\begin{thebibliography}{10}

\bibitem{Choptuik-1993-self-similarity}
M.~W. Choptuik, Physical Review Letters {\bf 70},  9  (1993).

\bibitem{Misner-1978-harmonic-maps}
C.~W. Misner, Physical Review D {\bf 18},  4510  (1978).

\bibitem{Eells-Lemaire-1978}
E. Eells and L. Lemaire, Bulletin of the London Mathematical Society {\bf 10},
  1  (1978).

\bibitem{Eells-Lemaire-1988}
E. Eells and L. Lemaire, Bulletin of the London Mathematical Society {\bf 20},
  385  (1988).

\bibitem{Bizon-1999-existence-of-self-similar-sigma-CSS-solutions}
P. \Bizon, Technical Report No.~math-ph/9910026, Jagellonian University,
  Krak\'ow, Poland (unpublished).

\bibitem{Bizon-Chmaj-Tabor-1999-sigma-3+1-evolution}
P. \Bizon, T. Chmaj, and Z. Tabor, Technical Report No.~4, Jagellonian
  University, Krak\'ow, Poland (unpublished).

\bibitem{Liebling-Hirschmann-Isenberg-1999-sigma-critical}
S.~L. Liebling, E.~W. Hirschmann, and J. Isenberg, Technical Report No.~8
  (unpublished).

\bibitem{Heusler-1996-No-hair-Theorems}
M. Heusler, Helvetica Physica Acta {\bf 69},  501  (1996).

\bibitem{Bizon-Chmaj-Tabor-1999-skyrme-3+1-crit-collapse}
P. \Bizon, T. Chmaj, and Z. Tabor, Physical Review D {\bf 59},  104003  (1999),
  (3 pages).

\bibitem{Goldwirth-Piran-1987-sssf-2+2}
D.~S. Goldwirth and T. Piran, Physical Review D {\bf 36},  3575  (1987).

\bibitem{Goldwirth-Ori-Piran-1989-in-Frontiers}
D.~S. Goldwirth, A. Ori, and T. Piran,  in {\em Frontiers in Numerical
  Relativity}, edited by C.~R. Evans, L.~S. Finn, and D.~W. Hobill (Cambridge
  University Press, Cambridge (UK), 1989), pp.\ 414--435, proceedings of the
  International Workshop on Numerical Relativity, University of Illinois at
  Urbana-Champaign (Urbana-Champaign, Illinois, USA), 9--13 May 1988.

\bibitem{Garfinkle-1995-sssf-2+2-self-similarity}
D. Garfinkle, Physical Review D {\bf 51},  5558  (1995).

\bibitem{GLPW-1996-sssf-3+1-and-2+2}
R. \Gomez{}, P. Laguna, P. Papadopoulos, and J. Winicour, Physical Review D
  {\bf 54},  4719  (1996).

\bibitem{Gomez-Winicour-1992-in-dInverno}
R. \Gomez{} and J. Winicour,  in {\em Approaches to Numerical Relativity},
  edited by R. d'Inverno (Cambridge University Press, Cambridge (UK), 1992),
  pp.\ 143--162, proceedings of the International Workshop on Numerical
  Relativity, Southampton University (Southampton, England), 16--20 December
  1991.

\bibitem{Gomez-Winicour-1992-sssf-2+2-asymptotics}
R. \Gomez{} and J. Winicour, Journal of Mathematical Physics {\bf 33},  1445
  (1992).

\bibitem{Gomez-Winicour-1992-sssf-2+2-numerical-methods}
R. \Gomez{} and J. Winicour, Journal of Computational Physics {\bf 98},  11
  (1992).

\bibitem{Wald}
R.~M. Wald, {\em General Relativity} (University of Chicago Press, Chicago,
  1984).

\bibitem{Hawking-Ellis}
S.~W. Hawking and G.~F.~R. Ellis, {\em The Large Scale Structure of Space-Time}
  (Cambridge University Press, Cambridge (UK), 1973).

\bibitem{Misner-Sharp-1964-Lagrangian-spherical-collapse}
C.~W. Misner and D.~H. Sharp, Physical Review B {\bf 136},  571  (1964).

\bibitem{Hayward-1994-mass-functions}
S.~A. Hayward, Physical Review D {\bf 49},  831  (1994).

\bibitem{Hayward-1996-Misner-Sharp-mass-function}
S.~A. Hayward, Physical Review D {\bf 53},  1938  (1996).

\bibitem{Koike-Hara-Adachi-1995-scaling-in-critical-collapse}
T. Koike, T. Hara, and S. Adachi, Physical Review Letters {\bf 74},  5170
  (1995).

\bibitem{Koike-Hara-Adachi-1996-scaling-in-critical-collapse}
T. Koike, T. Hara, and S. Adachi, Technical Report No.~gr-qc/9607010
  (unpublished).

\bibitem{Gundlach-1996-scaling-in-critical-collapse}
C. Gundlach, Physical Review D {\bf 55},  695  (1997).

\bibitem{Hod-Piran-1997-fine-structure-of-mass-scaling-law}
S. Hod and T. Piran, Physical Review D {\bf 55},  440  (1997).

\bibitem{Garfinkle-Duncan-1998-critical-curvature-scaling}
D. Garfinkle and G.~C. Duncan, Physical Review D {\bf 58},  064024  (1998), (4
  pages).

\bibitem{Liebling-inside-global-monopoles}
S.~L. Liebling, Physical Review D {\bf 60},  061502  (1999), (5 pages).

\bibitem{Bizon-Wasserman-2000-CSS-exists-for-nonzero-beta}
P. \Bizon{} and A. Wasserman, Technical Report No.~gr-qc/0006034, Jagellonian
  University, Krak\'ow, Poland (P.B.) and University of Michigan, U.S.A. (A.W.)
  (unpublished).

\bibitem{Courant-Friedrichs-Lewy-1928}
R. Courant, K. Friedrichs, and H. Lewy, Mathematische Annalen {\bf 100},  32
  (1928), (English translation in \cite{Courant-Friedrichs-Lewy-1967}).

\bibitem{Courant-Friedrichs-Lewy-1967}
R. Courant, K. Friedrichs, and H. Lewy, IBM Journal of Research and Development
  {\bf 11},  215  (1967), (English translation of
  \cite{Courant-Friedrichs-Lewy-1928}).

\bibitem{Hyman-1989-MOL-in-Buchler}
J.~M. Hyman,  in {\em NATO Advanced Research Workshop on the Numerical Modeling
  of Nonlinear Stellar Pulsations: Problems and Prospects}, edited by J.~R.
  Buchler (Kluwer, Dordrecht, 1989), pp.\ 215--237, also available as Los
  Alamos National Laboratories Report LA-UR 89-3136.

\bibitem{Choptuik-PhD}
M.~W. Choptuik, Ph.D. thesis, University of British Columbia, 1986.

\bibitem{Choptuik-1991-consistency}
M.~W. Choptuik, Physical Review D {\bf 44},  3124  (1991).

\bibitem{Choptuik-Goldwirth-Piran-1992-sssf-cmp-3+1-vs-2+2}
M.~W. Choptuik, D.~S. Goldwirth, and T. Piran, Classical and Quantum Gravity
  {\bf 9},  721  (1992).

\end{thebibliography}

%

\begin{table}
%
%
\begin{center}
\def\hleaders{ \leaders\hbox{\raise0.22em\hbox to 0.1em{\hrulefill}}\hfill\ }
\begin{tabular}{cccccccc}
   & \multicolumn{3}{c}{Initial Data Family~\eqref{eqn-Gaussian-initial-data}}
   & \multicolumn{4}{c}{Initial Data Family~\eqref{eqn-d4g-initial-data}}
                                                                        \\
   & \multicolumn{3}{c}{\hleaders{}Parameter is $A$\hleaders{}}
   & \multicolumn{4}{c}{\hleaders{}Parameter is $\sigma$\hleaders{}}
									\\
$\ccbeta$
        & $A^*$                 & $\Delta/2$    & $\gamma$
        & $A$   & $\sigma^*$    & $\Delta/2$    & $\gamma$
                                                                        \\
\hline 
\begin{tabular}{@{}r@{}l@{}}
  0&.18         \\
  0&.2          \\
  0&.225        \\
  0&.25         \\
  0&.3          \\
  0&.4          \\
  0&.5          \\
  1&            \\
  2&            \\
  5&            \\
 10&            \\
100&            
\end{tabular}
%
%
                & \begin{tabular}{@{}l@{}}
                  0.019\,523\,015   \\
                  0.018\,942\,512   \\
                  0.018\,241\,056   \\
                  0.017\,578\,042   \\
                  0.016\,392\,639   \\
                  0.014\,534\,866   \\
                  0.013\,167\,548   \\
                  0.009\,528\,975\,1  \\
                  0.006\,809\,778\,3  \\
                  0.004\,333\,205\,6  \\
                  0.003\,070\,144\,2  \\
                  0.000\,972\,589\,54 
                  \end{tabular}
                                & \begin{tabular}{@{}l@{}}
                                  0.5522        \\
                                  0.4367        \\
                                  0.3464        \\
                                  0.3043        \\
                                  0.2668        \\
                                  0.2452        \\
                                  0.2386        \\
                                  0.2314        \\
                                  0.2295        \\
                                  0.2304        \\
                                  0.2293        \\
                                  0.2302        
                                  \end{tabular}
                                                & \begin{tabular}{@{}l@{}}
                                                  0.1063      \\
                                                  0.1091     \\
                                                  0.1207     \\
                                                  0.1173     \\
                                                  0.1152     \\
                                                  0.1132     \\
                                                  0.1152     \\
                                                  0.1163     \\
                                                  0.1179     \\
                                                  0.1183      \\
                                                  0.1186     \\
                                                  0.1187     
                                                  \end{tabular}
%
%
        & \begin{tabular}{@{}l@{}}
          0.003         \\
          0.002         \\
          0.002         \\
          0.002         \\
          0.002         \\
          0.002         \\
          0.0015        \\
          0.0015        \\
          0.0010        \\
          0.0005        \\
          0.0005        \\
          0.0001        
          \end{tabular}
                & \begin{tabular}{@{}l@{}}
                  1.083\,153\,54        \\
                  0.615\,317\,49        \\
                  0.651\,519\,42        \\
                  0.688\,851\,73        \\
                  0.766\,003\,44        \\
                  0.929\,746\,89        \\
                  0.707\,335\,37        \\
                  1.210\,138\,07        \\
                  1.064\,744\,72        \\
                  0.734\,344\,76        \\
                  1.318\,800\,46        \\
                  0.631\,472\,58        
                  \end{tabular}
                                & \begin{tabular}{@{}l@{}}
                                  0.5478        \\
                                  0.4327        \\
                                  0.3472        \\
                                  0.3046        \\
                                  0.2675        \\
                                  0.2445        \\
                                  0.2386        \\
                                  0.2313        \\
                                  0.2305        \\
                                  0.2308        \\
                                  0.2312        \\
                                  0.2311        
                                  \end{tabular}
                                                & \begin{tabular}{@{}l@{}}
                                                  0.1028        \\
                                                  0.1150        \\
                                                  0.1169        \\
                                                  0.1173        \\
                                                  0.1146        \\
                                                  0.1139        \\
                                                  0.1130        \\
                                                  0.1155        \\
                                                  0.1167        \\
                                                  0.1178        \\
                                                  0.1182        \\
                                                  0.1182        
                                                  \end{tabular}
\end{tabular}
\end{center}
\caption{
	This table shows two families of near-critical initial data
	parameters for various coupling constants $\ccbeta$.
	For the Gaussian-like initial data
	family~\protect\eqref{eqn-Gaussian-initial-data}, we use the
	``amplitude''~$A$ as the parameter~$p$ (at a fixed
	``width''~$\sigma = 1$), with a numerical grid of
	\hbox{16\,000} grid points.
	For the ``derivative of 4th-power pseudo-Gaussian''
	initial data family~\protect\eqref{eqn-d4g-initial-data},
	we use the ``width''~$\sigma$ as the parameter~$p$
	(with different ``amplitudes''~$A$ for different
	coupling constants), with 8000 grid points.
	For each coupling constant and each family, the table also
	shows the $\max\,\, 2m/r$ echoing period $\Delta/2$ of the
	near-critical evolution, and the mass-scaling-law critical
	exponent $\gamma$ determined for the entire critical search.
	}
\label{tab-initial-data}
\end{table}


\begin{table}
\begin{center}
$$
\begin{array}{rccccc}
\multicolumn{1}{c}{\phantom{0}N}
        & p^*[N]        &       & \delta p^*
                                        &       & \text{ratio}          \\
\hline 
\begin{array}{r}
 1000   \\
 2000   \\
 4000   \\
 8000   \\
16\,000 
\end{array}
        & \begin{array}{l}
          0.013\,156\,008   \\
          0.013\,164\,618   \\
          0.013\,166\,841   \\
          0.013\,167\,405   \\
          0.013\,167\,548   
          \end{array}
                        & \begin{array}{l}
                          \smash{\rangle}\mathstrut     \\
                          \smash{\rangle}\mathstrut     \\
                          \smash{\rangle}\mathstrut     \\
                          \smash{\rangle}\mathstrut     \\
                          \end{array}
                                & \begin{array}{l}
                                  8.61 \times 10^{-6}   \\
                                  2.22 \times 10^{-6}   \\
                                  5.64 \times 10^{-7}   \\
                                  1.42 \times 10^{-7}   
                                  \end{array}
                                        & \begin{array}{l}
                                          \smash{\rangle}\mathstrut     \\
                                          \smash{\rangle}\mathstrut     \\
                                          \smash{\rangle}\mathstrut     \\
                                          \end{array}
                                                & \begin{array}{l}
                                                  3.87  \\
                                                  3.94  \\
                                                  3.97  
                                                  \end{array}
\end{array}
$$
\end{center}
\caption[Convergence of $p^*$ with Resolution]
        {
        This table shows the convergence of $p^*$ with the
	finite difference grid resolution~$N$, for the Gaussian-like
	data plotted in figure~\protect\ref{fig-pstar-convergence}.
        These evolutions used the same initial data parameters
        as given in table~\protect\ref{tab-initial-data}.
        The first two columns give the $p^*$ values for the various
        resolutions $N$.  The third column gives the differences
        $\delta p^* \equiv p^*[2N] - p^*[N]$ between consecutive
        $p^*$ values as the resolution is doubled, and the last
        column gives the ratios of consecutive differences.
        The values in the last column are very nearly equal
        to~4, showing second order convergence of $p^*$.
        }
\label{tab-pstar-convergence}
\end{table}

\end{document}